\documentclass[a4paper,UKenglish,cleveref, autoref, thm-restate]{lipics-v2021}

\pdfoutput=1 
\hideLIPIcs  

\newcounter{Problem}
\newtheorem{problem}[Problem]{Problem}

\usepackage{tikz}
\usetikzlibrary{positioning, arrows.meta, calc}
\usepackage{subcaption}

\nolinenumbers

\ifdefined\pdfprotrudechars\pdfprotrudechars=0\fi
\ifdefined\pdfadjustspacing\pdfadjustspacing=0\fi

\title{FBApro: A fast, simple linear transformation for diverse metabolic modeling tasks}

\author{Ariel Bruner}{Princeton University, Princeton, NJ, USA}{arielbruner@gmail.com}{https://orcid.org/0000-0003-3514-7769}{} 

\author{Mona Singh}{Princeton University, Princeton, NJ, USA}{mona@cs.princeton.edu}{https://orcid.org/0000-0001-8271-6026}{}

\authorrunning{A. Bruner and M. Singh}

\Copyright{Ariel Bruner and Mona Singh}



\category{} 

\relatedversion{} 
\relatedversiondetails[linktext={https://arxiv.org/abs/2601.14577}]{Preprint}{https://arxiv.org/abs/2601.14577}

\supplement{}

\supplementdetails[subcategory={Source Code}, cite={}, swhid={}]{Software}{https://github.com/Singh-Lab/FBApro}

\acknowledgements{This work was supported in part by the Ludwig Institute for Cancer Research, Princeton Branch and  NIH grant R01-CA208148. LLM tools were used for editing, proof reading, and generation of graphics.}

\EventEditors{Nadia El-Mabrouk and Fabio Vandin}
\EventNoEds{2}
\EventLongTitle{26th International Conference on Algorithms for Bioinformatics (WABI 2026)}
\EventShortTitle{WABI 2026}
\EventAcronym{WABI}
\EventYear{2026}
\EventDate{August 31--September 2, 2026}
\EventLocation{L'Aquila, Italy}
\EventLogo{}
\SeriesVolume{390}
\ArticleNo{28}

\begin{document}

\maketitle

\begin{abstract}
Constraint-based metabolic modeling is the predominant framework for simulating cellular metabolism. The central assumption of these models is that metabolism operates at a steady state, meaning that the production and consumption rates of each metabolite are balanced. This assumption imposes linear constraints on the fluxes of biochemical reactions. Flux Balance Analysis (FBA), a fundamental method in the field, is formulated as an optimization problem maximizing a cellular objective (e.g., growth) over the resulting linear subspace of steady state fluxes. Many other methods in the field are expressed either as a modification to FBA, or use FBA as a black box within an algorithm. Here, we propose a general alternative to optimization called FBApro. For any given vector of reference fluxes, FBApro finds the closest flux vector within the steady-state subspace, and accounts for both partially given reference fluxes and exact constraints on reactions. While FBApro is the solution to a quadratic program, we show that it can be implemented as a single linear operation using orthogonal projections to corresponding affine spaces and sets of linear equations. The overall approach is computationally efficient, does not require a cellular objective, and is easy to implement. We formally derive the closed-form expressions for FBApro and simpler variants, and validate it on both synthetic and real cancer cell line data.

\subparagraph{Code availability. }
The code implementing FBApro is available at \url{https://github.com/Singh-Lab/FBApro}. All code required to reproduce the figures in the paper is available, although the data used must be sourced separately. The repository also contains toy models and examples.
\end{abstract}

\newpage
\section{Introduction}

\begin{figure}[bthp]
    \centering
    \includegraphics[width=0.95\textwidth]{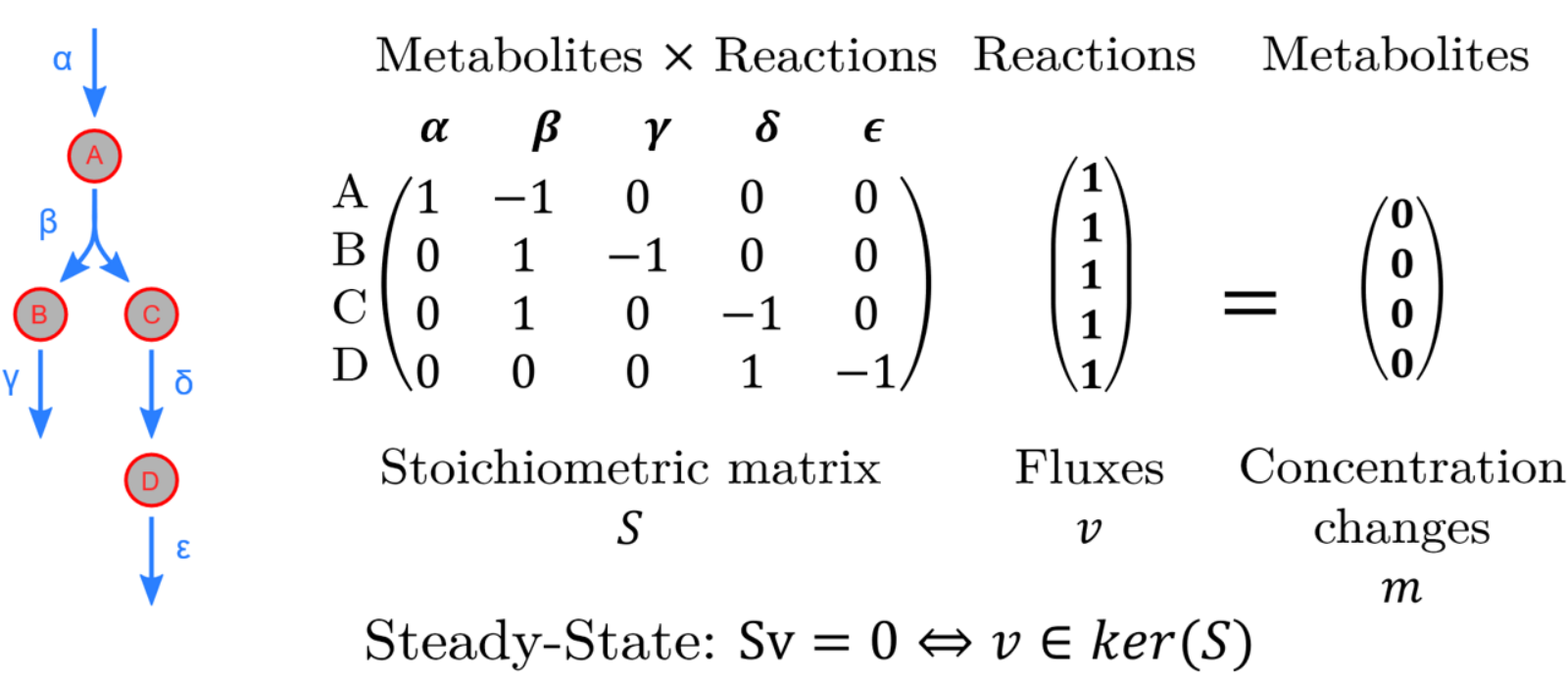}
    \caption{{\bf Left:} A metabolic model. Each node corresponds to a metabolite, and each hyperedge corresponds to a reaction. For example, reaction $\beta$ consumes metabolite $A$  and produces metabolites $B$ and $C$. {\bf Right:} The steady-state assumption in CBMM.  Each column in the stoichiometric matrix $S$ represents the metabolite changes induced by one reaction (e.g., see the column corresponding to the reaction $\beta$). Multiplying $S$ by the flux vector $v$, whose entries represent reaction rates, gives the net rate of change in metabolite concentrations. The steady state assumption in CBMM is that metabolite concentrations do not change over time; therefore feasible flux vectors satisfy $Sv=0$. }
    \label{fig:background}
\end{figure}

Cellular metabolism is a complex, tightly regulated process that underlies the behavior of biological systems. Modeling metabolism requires characterizing both the biochemical capabilities of organisms, as well as developing computational methods to simulate these processes under specific conditions. Constraint-Based Metabolic Mod\-eling (CBMM) has emerged as a widely-used framework for such simulations. It assumes a steady-state regime in which  metabolite concentrations remain constant and reaction fluxes must balance. Mathematically, the steady-state regime is expressed as the kernel of a stoichiometric matrix representing the metabolic model (see Figure \ref{fig:background}, and Appendix Section \ref{apxsec:fba} for more details).

CBMM has a long history, with theoretical foundations laid more than 40 years ago. Its most fundamental method is Flux Balance Analysis (FBA)~\cite{papoutsakis1985equations, orth2010flux, fell1986fat}, which formulates a linear program to capture steady-state flux constraints, individual reaction bounds and an organism-level objective. Large-scale sequencing efforts have enabled the con\-struc\-tion of genome-scale metabolic models for numerous organisms~\cite{bigg}, and FBA has been instrumental in predicting the metabolic phenotypes of unicellular organisms under various growth conditions~\cite{edwards2002metabolic, raman2009flux, kohlstedt2010metabolic, lewis2012constraining}. Extensions to FBA have also enabled FBA-based predictions of gene knockout effects (e.g., MoMA~\cite{segre2002analysis} and ROOM~\cite{shlomi2005regulatory}).

Modeling of multi-cellular organisms remains challenging due to heterogeneity across cell types, unclear metabolic objectives, and limited knowledge  of \emph{in vivo} metabolite availability.  These problems are even more pronounced in applications of CBMMs to cancer metabolism, where altered metabolism  plays a significant role \cite{cook2017genome}. 
The unique challenges of multicellular metabolic modeling, together with the increased availability of 'omics data, have led to the development of numerous methods integrating 'omics data with metabolic models. Such data can help tailor CBMMs to specific cell types, tissues, and disease contexts by identifying which metabolic genes and pathways are likely active, thereby helping to constrain the space of feasible fluxes and improving context-specific predictions. Gene expression (GE) data are the most widely available 'omics data, and many algorithms integrate GE with metabolic models to construct context-specific networks or infer context-specific reaction fluxes (e.g., GIMME \cite{becker2008context}, iMAT \cite{shlomi2008network}, MBA \cite{jerby2010computational}, INIT \cite{agren2012reconstruction}), including extensions to incorporate single-cell RNA-seq data (e.g., Compass \cite{wagner2021metabolic}). Most of these methods formulate optimization problems that incorporate both steady-state constraints and data fidelity terms, often resulting in large linear or mixed-integer programs, sometimes approximated by greedy heuristics. Scaling these approaches to large cohorts or single-cell datasets poses substantial computational challenges.

In parallel, advances in deep learning have enabled breakthrough advances in other areas of computational biology (e.g.,~\cite{jumper_alphafold_2021}), and highlight the value of domain-specific methods that admit closed-form differentiable im\-ple\-men\-ta\-tions that can be embedded directly into gradient-based machine learning models. However, to our knowledge, all available CBMM algorithms are based on solving optimization problems, particularly linear and mixed-integer linear programs, which are not easily integrated into end-to-end differentiable pipelines.

Here we present FBApro (short for FBA projection), a closed-form linear operation for solving the following general reference-flux agreement problem: given a reference flux vector $v$ for a metabolic model with $r$ reactions and steady-state space $ker(S)\subseteq \mathbb{R}^{r}$, a subset of high-confidence reactions $H\subseteq[r]$ and a disjoint subset of medium-confidence reactions $M\subseteq[r]$, find a steady-state flux vector that agrees with $v$ on $H$ and minimizes the $L_2$ distance to $v$ on $M$. This formulation models settings in which the reference vector $v$ combines heterogeneous sources of information: for example, we may have experimentally-derived flux measurements for reactions in $H$ (typically known for only a small number of reactions) and lower-confidence activity estimates for gene-associated reactions, inferred  from gene expression or other 'omics data  (not all reactions in metabolic models are associated with known genes). FBApro is presented as a projection onto an affine space determined by $S, M, H$ and the values of $v$ on $H$. Although this affine space depends on the reference flux input vector $v$,  we show that the resulting projection can nevertheless be expressed as a linear operator on $\mathbb{R}^r$ that is independent of~$v$.  Thus, for fixed $S$, $M$, and $H$, the projection matrix can be precomputed once for a given experimental setup and then applied by matrix multiplication to any number of reference flux vectors.  FBApro is therefore orders of magnitude faster than repeated applications of linear or quadratic programming based approaches when amortized over many inputs, while additionally being fully differentiable as a function of $v$. This amortized setting arises naturally in large-scale studies of human cancer cell lines or samples, where GE data are available across many biological contexts, direct flux measurements are sparse or unavailable, and the goal is to infer context-specific flux states across many samples.

FBApro is related to previous quadratic programming formulations, such as MoMA~\cite{segre2002analysis}, which minimizes the distance of a knockout model steady-state flux from a wild-type reference flux distribution, and the approach of~\cite{hackett2016systems} for fitting partially measured flux data. In contrast to these methods, FBApro is expressible as a simple orthogonal projection, at the cost of foregoing individual reaction bounds. To our knowledge, a projection-based method for solving the reference-flux agreement problem has not previously been formally described, proposed as an alternative to FBA-derived optimization methods, or validated on empirical data. Orthogonal projections have been used in the flux sampling module of the cobrapy metabolic modeling library~\cite{ebrahim2013cobrapy} for the special case of $M=[r]$, where they serve  as a computational step for mapping sampled points to the steady-state space.  In contrast, we show that projection can be used as the basis for solving more general reference-agreement problems involving both equality-constrained and medium-confidence reaction sets. We present simplified representations for FBApro for this case and the other special cases $H=\varnothing$ and $M\cup H=[r]$, with some cases admitting multiple equivalent expressions.

We illustrate the use of FBApro on an example ``toy'' model, and validate it on both simulated and cancer cell line data. On simulated data, we first compare the running time of FBApro to FBA, iMAT~\cite{shlomi2008network}, and MoMA~\cite{segre2002analysis}, and find that FBApro is orders of magnitude faster.  We compare FBApro to FBA as it is the \emph{de facto} method for metabolic modeling and has been used in various ways to incorporate GE values~\cite{covert2001regulation, kaste2023accurate, machado2014systematic}; to iMAT as it is a widely used method for integrating GE with metabolic models to uncover steady state fluxes that are ``consistent'' with the GE data; and to MoMA as it solves a similarly phrased problem to FBApro using the traditional optimization approach. The aforementioned methods also have an available implementation written in Python. We then consider simulations where exact, noisy and missing data are derived from steady state fluxes, and demonstrate that FBApro is competitive with previous approaches in recapitulating the known steady state fluxes, while requiring a fraction of the running time. Finally, we consider cancer cell-line GE and reaction flux data to show that FBApro surpasses or has competitive performance in imputing unmeasured fluxes and mapping GE data to the steady-state space while predicting corresponding unseen data.  Overall, our results suggest that FBApro is a flexible, modular and effective approach for high-throughput and large-scale metabolic modeling. 

\subparagraph{Organization of remainder of the paper.}
In Section~\ref{sec:problems}, we give formal definitions of the variants of FBApro and state the theorems that allow us to derive the algorithms underlying FBApro. In Section \ref{sec:proofs}, we prove the theorems and derive the expressions for variants. In Section~\ref{sec:usecases}, we show how FBApro can be utilized to integrate data into a model, using a ``toy'' metabolic model.  In Section~\ref{sec:validation}, we demonstrate the performance of FBApro on simulated and real empirical data, and give some concluding remarks in Section~\ref{sec:discussion}.

\section{Problem Statements}
\label{sec:problems}

\subsection{Definitions and Notations }

Let $m,r\in \mathbb{N}$ be the number of metabolites and reactions in a metabolic model, and $S\in \mathbb{R}^{m\times r}$ be the stoichiometric matrix of the model, such that a unit of flux in reaction $j$ produces $S_{i,j}$ units of metabolite $i$ (or consumes, if negative). The steady-state space of the model is the set of flux assignments for which each metabolite has zero net production;  i.e., $ker(S)=\{v\in \mathbb{R}^r|Sv=0\}$. As an alternative representation for the steady-state space, let $l=dim(ker(S))$ and let $A\in \mathbb{R}^{r\times l}$ be a matrix whose columns form an orthonormal basis of $ker(S)$, so that $col(A)=ker(S)$ and $A^TA=I$, computable from $S$ using a Singular Value Decomposition (SVD). 

Our methods use the Moore-Penrose inverse for projections and solutions for systems of linear equations. For a matrix $X$, let $X^+$ denote its Moore-Penrose inverse, or pseudoinverse (also  computable from $X$ using SVD). We also make use of vector restrictions, and associated restriction matrices: for a vector $v$ and index set $Q=\{q_1,\ldots, q_{|Q|}\}$, denote by $v_Q=(v_{q_1},\ldots, v_{q_{|Q|}})^T$ the restriction of $v$ to $Q$, and let $P_Q\in \{0,1\}^{|Q|\times r}$ be the restriction matrix to $Q$,
\begin{align*}
(P_Q)_{i,j}=
\begin{cases}
1 & j=q_i\\
0 & \text{ otherwise,}
\end{cases}
\end{align*}
such that $\forall v.P_Qv=v_Q$. Similarly, for a matrix $X\in \mathbb{R}^{a\times b}$ and index sets $\mathcal{I}\subseteq [a], \mathcal{J}\subseteq [b]$, we denote the submatrix restricted to these sets by $X_{\mathcal{I},\mathcal{J}}=\{X_{i,j}|i\in \mathcal{I},j\in \mathcal{J}\}$. Denote by $M,H\subseteq [r]$ the indices of medium and high confidence reactions, respectively, with an arbitrary order over the sets. Let $B\in\mathbb{R}^{r\times (r-|H|)}$ be the matrix whose columns are the standard basis vectors $\{e_i|i\notin H\}$ of $\mathbb{R}^r$, spanning all flux vectors with the reactions of $H$ carrying no flux. 

\subsection{FBApro: General case }
The main problem we solve is the general problem of finding a steady-state flux $x\in \mathbb{R}^r$ given $S$ approximating a given reference flux $v\in \mathbb{R}^r$ for reactions with indices in $M\subseteq [r]$ and equaling $v$ for reactions with indices in $H\subseteq [r]$, with $H\cap M = \varnothing$.  Our problem can be formulated as:

\begin{problem}\label{opt:general}
\begin{align*}
FBApro(S,M,H,v)=    argmin_{x\in \mathbb{R}^{r}}\quad & \left \lVert x_M-v_M \right\rVert _2^2\\
    \text{subject to} \quad & Sx=0\\
    \quad & x_{H}=v_H,
\end{align*}
\end{problem}

where $argmin$ refers to any solution minimizing the distance. An algorithm that solves FBApro in this most general form derives from the following theorem:
\begin{theorem}\label{thm:general}

With $P\mathrel{\mathop:}= P_{M}$ the restriction matrix to $M$, and $B$ spanning all fluxes in reactions not in $H$, define 
\begin{align*}
D=&AA^T + BB^T, &C=&[BB^TD^+A:AA^TD^+B], &E=C(PC)^+P,
\end{align*}
with $:$ denoting concatenation of columns. Then, if a solution to Problem \ref{opt:general} exists, 
\[
FBApro(S,M,H, v)=\left(E(I_r-AA^TD^+)+AA^TD^+\right)v.
\]
\end{theorem}

The proof of this Theorem is given in Subsection~\ref{subsec:full_proof}. In practice, this theorem shows that, given $v$, FBApro($v$) can be simply expressed as a multiplication of $v$ by a precomputed matrix\footnote{We will sometimes refer to FBApro as FBAproFull to differentiate it from other simpler versions of the algorithm.}. Computing this matrix requires a constant number of matrix multiplication, addition and concatenation operations, as well as a constant amount of SVD computations, all for matrices of $O(max\{r,m\})$ rows and columns. Typical metabolic models have numbers of reactions and metabolites on the same order, and so assuming $m\in \Theta(r)$ the entire preprocessing has time complexity $O(r^3)$, and theoretically lower if fast matrix multiplication is considered. Each subsequent application of FBApro for different reference value vectors has a time complexity of $O(r^2)$.

Linear programs like FBA are typically solved with the simplex method, which has a worst-case exponential complexity but is polynomial in practice, and provably polynomial methods exist admitting a complexity of $O(r^3)$, similarly improved by fast matrix multiplication. Convex quadratic programs like MoMA also have a cubic complexity, in contrast to mixed-integer programs like iMAT which are NP-hard. Note that unlike FBApro, there is no improvement in amortized time complexity gained by solving these programs for multiple samples.

\subsection{FBAproBasic: Noisy Flux Estimates for All Reactions}
Before proving our main result, we also consider special cases of FBApro, the simplest of which is formalized next. 

\begin{problem}\label{opt:basic}
\begin{align*}
   FBAproBasic(S,v)= argmin_{x\in \mathbb{R}^{r}}\quad & \left \lVert x-v \right\rVert _2^2\\
    \text{subject to} \quad & Sx=0,
\end{align*}
\end{problem}

corresponding to $M=[r]$. This formulation is similar in spirit to the optimization problem posed by MoMA~\cite{segre2002analysis}, differing in the exclusion of reaction bounds $l\leq v\leq u$, and the typical MoMA use case of simulating knockout behavior with the constraint $v_i=0$ for some reaction $i$. Since a knockout can also be represented either by setting $l_i=u_i=0$ or by considering a stoichiometric matrix with column $i$ removed, only the reaction bounds meaningfully differentiate the two problems. However, unlike MoMA, FBAproBasic admits a closed-form solution using an orthogonal projection, as described in the next theorem:

\begin{theorem}\label{thm:basic}
    \begin{equation*}
        FBAproBasic(S, v)=AA^Tv=(I-S^+S)v.
    \end{equation*}
\end{theorem}

FBAproBasic is equivalent to an orthogonal projection to a linear subspace, a standard linear algebra operation. Nevertheless, for completeness and as a ``warm up'' for the other proofs, we include in Appendix Section \ref{apxsec:basic_proof} a standalone proof for Theorem~\ref{thm:basic}, relying only on basic linear algebra and the properties of the pseudoinverse.

\subsection{FBAproPartial: Handling Unknown Flux Values}
The following case corresponding to $H=\varnothing$ is not directly represented as an orthogonal projection, and to our knowledge has not been studied before:

\begin{problem}\label{opt:partial}
\begin{align*}
   FBAproPartial(S,M,v)= argmin_{x\in \mathbb{R}^{r}}\quad & \left \lVert x_M-v_M \right\rVert _2^2\\
    \text{subject to} \quad & Sx=0,
    \end{align*}
\end{problem}

for which we show:

\begin{theorem}\label{thm:partial}
    \begin{equation*}
    FBAproPartial(S, M, v)=A(P_MA)^+P_Mv.
    \end{equation*}
\end{theorem}
Thus, as with general FBApro and the simple FBAproBasic, the algorithm for finding the flux vector satisfying our constraints corresponds to precomputation of a matrix based on the metabolic model and knowledge of which reactions mid-confidence, followed by a simple matrix multiplication involving the input flux vector.

\subsection{FBAproFixed: Handling Exact Flux Values}

The case corresponding to $M\cup H=[r]$ leads to the following problem:

\begin{problem}\label{opt:fixed}
\begin{align*}
    FBAproFixed(S,H,v)=argmin_{x\in \mathbb{R}^{r}}\quad & \left \lVert x-v \right\rVert _2^2\\
    \text{subject to} \quad & Sx=0\\
    \quad & x_{H}=v_H,
\end{align*}
\end{problem}
This problem can be thought of as an orthogonal projection to the affine space $(v+col(B))\cap ker(S)$, if a solution exists and the intersection is not empty. We first derive an expression for FBAproFixed using basic linear algebra:

\begin{theorem}\label{thm:fixed1}

Define $H^c=[r]\setminus H$ and define $Q\in \mathbb{R}^{r\times r}$ such that its restriction submatrices are

\begin{align*}
    &Q_{H,H}&=&I_{|H|}\\
    &Q_{H, H^c}&=&0\\
    &Q_{H^c,H}&=&-(S_{[m],H^c})^+S_{[m],H}\\
    &Q_{H^c,H^c}&=&\left(I_{|H^c|}-(S_{[m],H^c})^+S_{[m],H^c}\right).
\end{align*}

If a solution to Problem \ref{opt:fixed} exists, then $FBAproFixed(S, H, v)=Qv$.
\end{theorem} 
We prove this derivation in the Appendix Section \ref{apxsec:fixed_proof}. Before proving our main theorem, we draw on \cite{Projections} to derive the following alternative representation:

\begin{theorem}\label{thm:fixed2}
As previously defined, let \begin{align*}
D=&AA^T + BB^T, &C=&[BB^TD^+A:AA^TD^+B],
\end{align*}

If a solution to Problem \ref{opt:fixed} exists, then

    \begin{equation*}
        FBAproFixed(S, H, v)=\left(CC^+\left(I_{r}-AA^TD^+\right)+AA^TD^+\right)v.
    \end{equation*}

\end{theorem}

The proof is given in Subsection \ref{subsec:fixed_proof}, and is followed by the proof of the main theorem in Subsection \ref{subsec:full_proof}.

\section{Deriving FBApro}\label{sec:proofs}

Before proving Theorem~\ref{thm:general} in Section~\ref{subsec:full_proof}, which gives the formula for solving FBApro in its most general form, we first derive formulas for two simpler cases. In Section~\ref{subsec:partial_proof}, we consider FBAproPartial, where the reference flux contains medium-confidence, approximate reaction values that we aim to match as closely as possible, as well as reactions for which no reference information is available. In Section~\ref{subsec:fixed_proof}, we consider FBAproFixed, where the reference flux contains high-confidence reaction values that must be maintained exactly, with the remaining values optimized over. The proof techniques developed for these two subproblems are then combined to derive the general FBApro solution. 

\subsection{FBAproPartial}\label{subsec:partial_proof}

Recall that $FBAproBasic(S, v) = AA^Tv=(I-S^+S)v$. While FBAproBasic can alternatively be computed using either $S$ or $A$, representing the steady-state space as $col(A)$ allows for a straightforward representation of the steady-state restrictions using the restriction matrix $P\mathrel{\mathop:}= P_{M}$, yielding our single representation for FBAproPartial: 

\begin{proof}[Proof of Theorem \ref{thm:partial}]
Recall that $\forall x. Px=x_M$. Then $\{x_M|x\in col(A)\}=col(PA)$ and so 

\[\min_{x\in col(A)}\lVert v_M-x_M\rVert_2^2=\min_{y\in col(PA)}\lVert Pv-y \rVert_2^2.\] Since $col(PA)$ is a linear space spanned by $PA$, we know from the proof of Theorem \ref{thm:basic} that a distance minimizing $y$ in the space can be expressed as $y=(PA)(PA)^+(Pv)$. Finally, we want an $x\in\operatorname{col}(A)$ satisfying $Px=y$, but we already have the expression $y=P(A(PA)^+Pv)$, yielding $x=A(PA)^+Pv$ as a solution, as needed. \end{proof}

\subsection{FBAproFixed}\label{subsec:fixed_proof}

While FBAproFixed admits an expression relying only on basic linear algebra, derived in Appendix Section \ref{apxsec:fixed_proof}, it does not generalize well to the general problem. For this, we rely on a closed-form expression for an orthogonal projection to the intersection of two affine space given in \cite[Theorem~4.1]{Projections}, restated here:

\begin{theorem}\label{thm:projections}
    If $T:=(a+col(A))\cap (b+col(B))$ is not empty, and with $D:=AA^*+BB^*$, then $T=c+col(C)$, where 

    \[
    c=AA^*D^+(b-a)+a
    \]

and, denoting column concatenation with $:$,  

\[
C = \left[BB^*D^+A:AA^*D^+B \right].
\]
\end{theorem}

We can now derive the affine intersection formulation of FBAproFixed:  

\begin{proof}[Proof of Theorem \ref{thm:fixed2}]
    
Recall that $A$ is an orthogonal basis spanning the steady-state space, and that $B$ is defined such that $\forall x. x+col(B)=\{y|y_H=x_H\}$. Then the space we seek an orthogonal projection to is $col(A)\cap (v+col(B))$. Following Theorem \ref{thm:projections}, and given $A,B$ are real, we define 
    
    \begin{align*}
     D&=AA^T+BB^T\\
     c&=AA^TD^+v\\
     C&=[BB^TD^+A:AA^TD^+B],
    \end{align*}

and the projection of $v$ to $c+col(C)=col(A)\cap (v+col(B))$ is 

\[
CC^+(v-c)+c=CC^+(v-AA^TD^+v)+AA^TD^+v=\left(CC^+(I_{r}-AA^TD^+)+AA^TD^+\right)v.
\]
\end{proof}

\subsection{FBApro}\label{subsec:full_proof}

To derive FBApro, we combine the restriction method used for FBAproPartial with the affine intersection representation of FBAproFixed:

\begin{proof}[Proof of Theorem \ref{thm:general}]

With $c$ and $C$ as before, and with $P\mathrel{\mathop:}=P_{M}$, we want to find $x$ solving 

\begin{align*}
     \min_{x\in \mathbb{R}^{r}}\quad & \left \lVert x_M-v_M \right\rVert _2^2\\
    \text{subject to} \quad & Sx=0 \\
    \quad & x_{H}=v_H,&&
\end{align*}
equivalently expressed as
\[
\min_{x\in c + col(C)} \lVert v_M-x_M \rVert_2^2 = \min_{y\in Pc + col(PC)}\lVert Pv-y \rVert_2^2,
\]
and solved by an orthogonal projection of $Pv$ to $Pc+col(PC)$. Denoting $E=C(PC)^+P$, this is given by

\begin{align*}
&PC(PC)^+(Pv-Pc)+Pc\\
&\quad=PE(v-AA^TD^+v)+PAA^TD^+v\\
&\quad=\left(PE(I_{r}-AA^TD^+)+PAA^TD^+\right)v\\
&\quad=P \left(E(I_{r}-AA^TD^+)+AA^TD^+\right)v.    
\end{align*}

Given a projection of $Pv$ to $Pc+col(PC)$ denoted $y$, we now seek $x$ such that $Px=y$. As in FBAproPartial, our expression yields such a solution since it is a right product of $P$, and thus,

\[
FBApro(S, M, H, v)=\left(E(I_r-AA^TD^+)+AA^TD^+\right)v.
\]

\end{proof}

\section{Use Cases}\label{sec:usecases}

In this section, we illustrate the use of FBApro and its variants on a toy metabolic network, demonstrating their behavior on data imputation. 

\subsection{Toy Model}\label{sec:toymodel}

Consider the branched pathway in Figure~\ref{fig:network}, loosely analogous to a metabolic branch point such as the splitting of a precursor between two biosynthetic routes with different stoichiometric yields. The network has three metabolites $A, B, C$, five reactions, and the stoichiometric matrix

\[
S=\begin{pmatrix} 1 & -1 & -1 & 0 & 0 \\ 0 & 1 & 0 & -1 & 0 \\ 0 & 0 & 2 & 0 & -1 \end{pmatrix},
\]
where reaction $v_1$ imports $A$ from the environment, $v_2$ converts one unit of $A$ into one unit of $B$, $v_3$ converts one unit of $A$ into two units of $C$, and $v_4, v_5$ export $B$ and $C$ respectively. The steady-state space is two-dimensional, parametrized by $t,s\in\mathbb{R}$ as 

\[ker(S)=\{(t+s,\; t,\; s,\; t,\; 2s)\mid t,s\in \mathbb{R}\}.\]
Note that the network definition does not explicitly include bounds on reactions, or the requirement that $t, s\geq 0$. 

\begin{figure}[t]
\centering
\begin{tikzpicture}[
    metab/.style={circle, draw, thick, minimum size=8mm, inner sep=0pt},
    rxn/.style={-{Stealth[length=3mm]}, thick},
    lbl/.style={font=\footnotesize, fill=white, inner sep=1pt}
]
\node[metab] (A) at (0,0) {$A$};
\node[metab] (B) at (3,1.2) {$B$};
\node[metab] (C) at (3,-1.2) {$C$};

\draw[rxn] (-2,0) -- node[lbl, above=0.1cm] {$v_1$} (A);
\draw[rxn] (A) -- node[lbl, above left=0.1cm] {$v_2: 1{\to}1$} (B);
\draw[rxn] (A) -- node[lbl, below left=0.1cm] {$v_3: 1{\to}2$} (C);
\draw[rxn] (B) -- node[lbl, above=0.1cm] {$v_4$} (5,1.2);
\draw[rxn] (C) -- node[lbl, above=0.1cm] {$v_5$} (5,-1.2);
\end{tikzpicture}
\caption{Toy branched pathway. One unit of $A$ is converted to either one unit of $B$ (via reaction $v_2$) or two units of $C$ (via reaction $v_3$).}\label{fig:network}
\end{figure}

\subsection{Data Imputation on Toy Model}\label{sec:imputation}

Suppose the ground-truth steady-state flux is $v^*=(7, 4, 3, 4, 6)$, corresponding to $t=4, s=3$. We observe a partial and noisy reference: $v_1=7$ is measured with high confidence, $v_3\approx 2.2$ and $v_4\approx 3.6$ are measured with moderate noise, while $v_2$ and $v_5$ are unobserved. We assign $H=\{1\}$, $M=\{3,4\}$ for FBApro (FBAproFull henceforth, for readability), and set the reference (input) vector $\tilde{v}=(7, 0, 2.2, 3.6, 0)$ where entries for unobserved reactions are set to zero. For FBAproPartial, we set $M=\{1,3,4\}$ and treat the first reaction as a noisy input. For FBAproFixed, we set $H=\{1\}$ and $M=\{2,3,4,5\}$, and for FBAproBasic $M=[r]$. Table~\ref{tab:imputation} shows the output of each FBApro variant.

The results illustrate the value of the three-level confidence structure. FBApro correctly treats the high-confidence measurement as a hard constraint, fits the noisy measurements as well as the steady-state constraint allows, and does not penalize deviations on the unobserved reactions. Each special case loses one of these advantages: FBAproPartial treats the measurement of $v_1$ as noisy, losing information, FBAproFixed assumes unmeasured fluxes should be minimized in absolute value, and FBAproBasic suffers from both effects simultaneously. While this order of performance is expected, we note that different orders are possible depending on different factors, such as the number of unmeasured reactions, or a positive or negative bias in noisy values.

\begin{table}[h!]
\caption{Data imputation results on the toy model. The reference $\tilde{v}$ uses zero for unobserved reactions (marked with ``u''). Medium-confidence or approximate values are marked with $\approx$. Distances are $L_2$ distances of the output $\hat{v}$ to the ground truth $v^*$.}\label{tab:imputation}
\centering
\begin{tabular}{l c c c c c c}
\hline
 & $v_1$ & $v_2$ & $v_3$ & $v_4$ & $v_5$ & $\lVert \hat{v}-v^*\rVert$ \\
\hline
Ground truth $v^*$ & 7& 4& 3& 4& 6 & --- \\
Reference (input) $\tilde{v}$& 7& 0 (u)& $\approx 2.2$& $\approx 3.6$& 0 (u)& --- \\
\hline
FBAproFull $\hat{v}$& 7& 4.2& 2.8& 4.2& 5.6& 0.53\\
FBAproPartial $\hat{v}$ & 6.6& 4& 2.6& 4& 5.2& 0.98\\
FBAproFixed $\hat{v}$ & 7& 5.2& 1.8& 5.2& 3.6& 3.17\\
FBAproBasic $\hat{v}$ & 4.2& 3.2& 1& 3.2& 2& 5.4\\
\hline
\end{tabular}
\end{table}

While this imputation problem is natural for FBApro, it does not have a natural representation using FBA and most derived methods. FBA can encode noisy observations only as flux bounds (e.g.\ $v_3\in [2.5-\epsilon, 2.5+\epsilon]$), requiring an explicit noise model and an appropriate choice of $\epsilon$. iMAT reduces the quantitative reference to a ternary classification of reactions as ON, OFF or unconstrained, discarding the numerical signal. MoMA, being similar in formulation, can recreate the performance of FBAproBasic, and analogues of FBAproPartial and FBAproFixed could be formulated as quadratic programs with the appropriate modifications. However, to our knowledge, partial and fixed variants of MoMA have never been introduced, and would regardless require the optimization of a quadratic program (or a relaxed linear program).

\section{Empirical Validation}\label{sec:validation}

\subsection{Methods and Their Inputs}\label{subsec:methods}

FBApro and the benchmark methods differ in the inputs they require and in how they use the available reference information. In all experiments, the underlying metabolic model provides a stoichiometric matrix and default reaction bounds. The methods then differ in whether they additionally require an objective function, a reference vector, or reaction confidence sets. All methods other than FBApro and its variants are run via their implementation in cobrapy~\cite{ebrahim2013cobrapy}.

\subparagraph{FBApro.}
FBApro variants take as input a metabolic model and a reference flux vector, together with reaction confidence information. FBAproBasic uses only the reference flux and the model, without distinguishing among reactions by confidence. FBAproPartial additionally uses a set of medium-confidence reactions, denoted $M$, whose reference values are treated as approximate measurements to be matched as closely as possible. FBAproFixed uses a set of high-confidence reactions, denoted $H$, whose reference values are enforced exactly, while optimizing over the remaining reactions. FBAproFull uses both sets: medium-confidence reactions $M$ and high-confidence reactions $H$.

By running special cases of FBApro on data, we effectively ignore some of the information on reaction confidences. Let $M_0,H_0$ be the confidences we derived from the data. FBAproBasic uses $M=[r]$, ignoring the different confidences completely. FBAproPartial uses $M=M_0\cup H_0$, interpreting any reaction with associated data as medium confidence, and FBAproFixed uses $H=H_0,M=[r]\setminus H_0$, interpreting all reactions without high confidence as medium confidence.

\subparagraph{FBA.}
FBA takes as input a metabolic model, specifically a stoichiometric matrix and reaction bounds, as well as a linear objective function, and computes a steady-state flux vector within the reaction bounds maximizing the objective (see Appendix Section \ref{apxsec:fba} for details). In all experiments, we used each model's default objective, typically the biomass reaction. To apply FBA in settings with gene expression or flux measurements, we used the reference values to modify the model's reaction bounds, allowing the measurements to constrain the feasible flux space. For a reaction with reference value $x$ in a particular sample, and with an assumed gap $l$, we set the reaction bounds to $[x-l,x+l]$. In the synthetic data experiments, we used $l=10^{-5}$ for fixed reactions, and $l=|X|+10^{-5}$ for noisy reactions (matching the model we used to generate noise). For real data, we used $l=|X|$. For unmeasured reactions, we used the model's bounds.

\subparagraph{iMAT.}
iMAT takes a metabolic model and a reference vector as input, which it interprets as activity measures derived from gene expression (GE) data. It then finds a steady-state flux vector with high agreement to the reference vector under steady-state constraints (see Appendix Section \ref{apxsec:imatmoma}).  As with FBA, we use the reference values to define reaction bounds for each sample. 

\subparagraph{MoMA.}
MoMA takes as input a metabolic model and a reference flux vector, which it interprets as fluxes, and finds via quadratic programming the feasible closest steady-state flux vector to that reference under steady-state constraints (see Appendix Section \ref{apxsec:imatmoma}). Due to technical issues with the cobrapy implementation of MoMA when using sample-specific bounds, we use each model's default reaction bounds for MoMA in all samples.

\subsection{Timing Methods}

To measure the running time of FBApro and several benchmark methods, we retrieved five metabolic models for representative organisms from the BiGG Models repository \cite{bigg}: a small \emph{E. coli} core metabolism model (e\_coli\_core),  a genome scale model for \emph{E. coli K-12}  (iML1515), a yeast model (iND750), a mouse model (iMM1415) and a human model (RECON1). For each model, we first compute a spanning basis for its steady-state flux space and sample 50 steady-state flux vectors by taking linear combinations of the basis vectors with coefficients uniformly sampled from $[0, 1]$. We then select uniformly at random 10\% of the reactions to serve as known reactions, and 50\% of the model's reactions as unmeasured reactions. 

To generate reference fluxes from the sampled steady-state flux vectors, we retain the values of the known reactions, set the unmeasured reaction values to zero, and add multiplicative noise to the remaining reactions as follows. For a true value $x$, we sample the corresponding noisy value $\tilde{x}$ uniformly from $[x-|x|,x+|x|]$. As described in Subsection \ref{subsec:methods}, for methods requiring bounds as inputs, we use the model's default bounds for unknown reactions, $\tilde{x}\pm \epsilon$ with $\epsilon = 10^{-5}$ for known reactions, and for noisy reactions, $[\tilde{x}-|\tilde{x}|-\epsilon, \tilde{x}+|\tilde{x}|+\epsilon]$. Note that these bounds do not necessarily contain the true steady-state fluxes. More generally, deriving bounds from empirical measurements that are both feasible and informative is a challenging problem.

We measure the running time of FBApro and the benchmark methods on these simulated fluxes. For FBApro variants, we separately measure the setup time and application time. The setup time consists of computing the projection matrix used by each method, whereas the application time is the time required to apply this precomputed projection to a new reference flux. This distinction is important because the setup cost is paid once for a fixed model and measurement structure (i.e., which reaction measurements are fixed, noisy, or unknown), while the projection can then be reused across many samples. We note that while FBA, iMAT and MoMA have associated setup steps, notably constructing the associated linear or mixed-integer program, these are not explicitly exposed through cobrapy. Our measurements for these methods capture the average running time of a single method call. Some of the setup time, specifically the base steady-state linear program, is done by cobrapy at model load time and is thus not captured by our measurements.

For FBApro, we separately measure the setup and running time using CPU and GPU. We also measure the amortized application time per sample when the precomputed projection is applied to all 50 samples as a single input matrix. All time measurements were taken on a node with 1 CPU core, 1 NVIDIA MIG GPU, 32GB of CPU memory and 10 GB of GPU memory, allocated from the Della cluster of Princeton Research Computing. 

\begin{figure}[bhtp]
    \centering
    \includegraphics[width=0.95\textwidth]{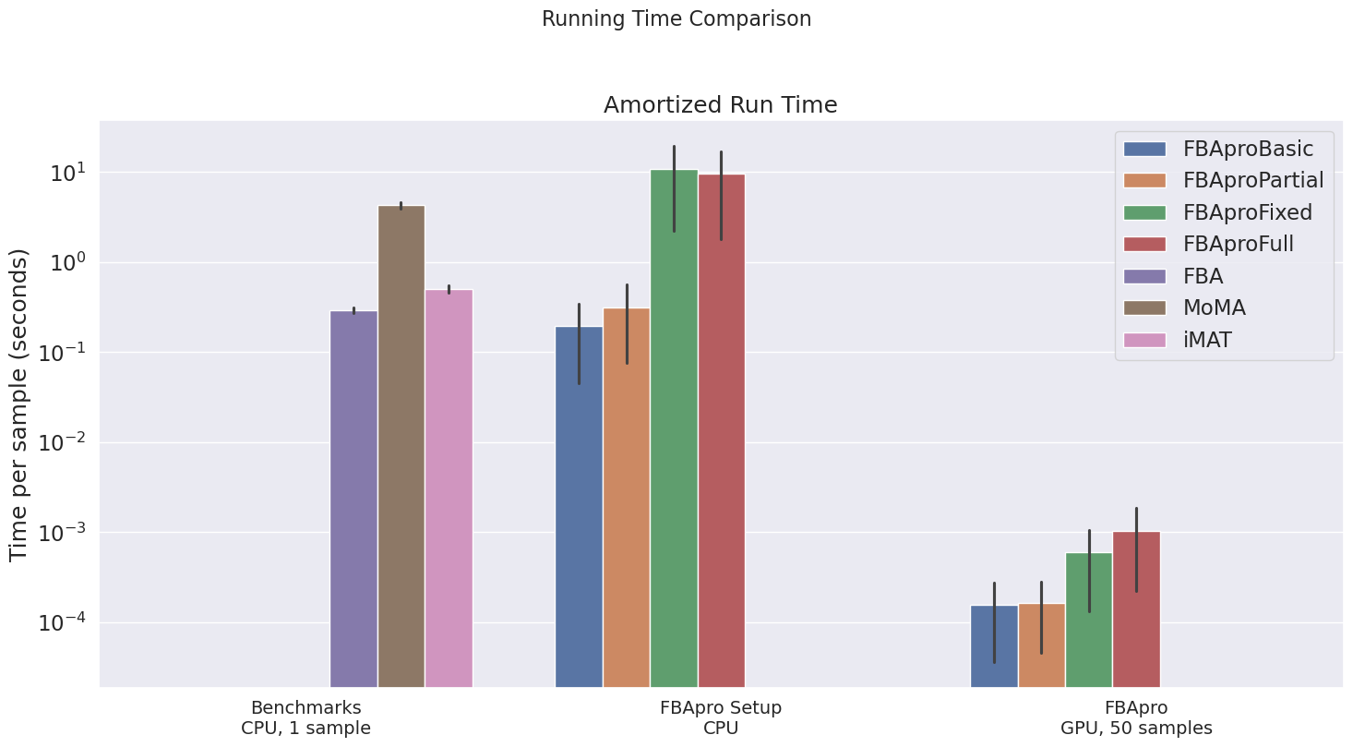}
    \caption{Timing results, in seconds, on a log scale, averaged over five models and 50 samples per model. For FBApro variants, setup time on CPU is measured separately, and amortized runtime per sample on GPU is measured with the inputs given as a single matrix. FBApro and its variants are orders of magnitude faster than the previous approaches based on FBA, MoMA, and iMAT. Error bars represent a 95\% confidence interval.}
    \label{fig:timing}
\end{figure}

For reasons relating to the numerical stability of underlying linear algebra drivers and their availability on CPU and GPU, we recommend performing the FBApro setup on CPU, and applying the resulting projection in batches on GPU. Figure \ref{fig:timing} shows the corresponding running time results, with the full comparison available in Appendix Figure \ref{fig:apx_timing}. A single application of any of the FBApro projection methods is orders of magnitude faster than a single application of any of the benchmark methods. This speedup is especially pronounced in the batched setting, where the fixed setup cost is amortized across many reference fluxes.

The main computational consideration is that the setup time for FBApro variants, particularly the Fixed and Full versions, is comparable to the running time of MoMA. However, this is a fixed cost incurred only once for a given metabolic model and measurement configuration, after which each additional reference flux can be processed very quickly. Thus, FBApro is particularly advantageous in settings involving repeated inferences over many samples measured on the same reactions set. These situations naturally arise, for example, when trying to infer fluxes from gene expression data for large numbers of tumor samples (such as from TCGA~\cite{TCGA}); in this case, across samples, the same reactions with known associated genes have estimates of activity that can be used to constrain flux estimates. For the simplest variant, FBAproBasic, the precomputation is dependent only on the metabolic model, and can be done on model load time, as currently implemented in cobrapy for the setup of standard FBA.

\subsection{Inferring Hidden Synthetic Data Fluxes}

We next evaluate the ability of FBApro and its variants to infer hidden fluxes with a similar experimental setup. We generated synthetic data on the human metabolic model RECON1. First, 2\% of the reactions are randomly selected as known, a random 85\% are treated as unmeasured, and the remaining 13\% are provided with a multiplicative noise of 100\%. 25 steady-state samples are then generated, and processed to reference flux input vectors. After running each method on these data, for each noisy reaction, we compute the Spearman correlation between the fluxes returned by each method and the corresponding ground truth fluxes across samples. We similarly compute the Spearman correlation across the same reactions for each sample. For any method, if the predictions are constant across the computed axis, we assign a correlation of zero. We repeat the experiment for 10 independent selections of known and unmeasured reactions, and aggregate the results, resulting in correlations over 250 samples in total. 

\begin{figure}[htbp]
    \centering
    \begin{subfigure}[b]{0.475\textwidth}
        \centering
        \includegraphics[height=4.5cm]{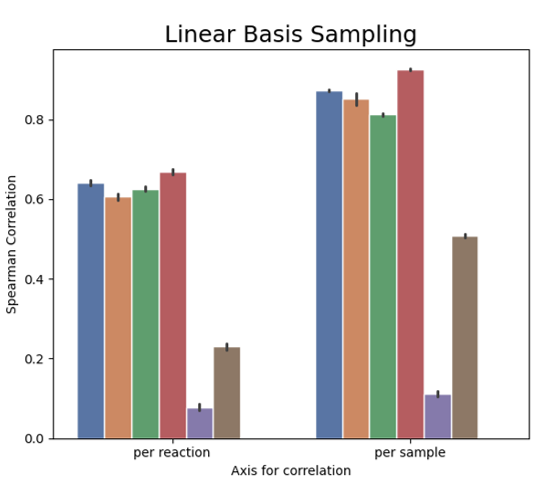}
        \caption{Spearman correlations between ground truth fluxes and method predictions, data sampled using random combinations of a spanning basis.}
        \label{fig:synthetic_correlations_a}
    \end{subfigure}
    \hfill
    \begin{subfigure}[b]{0.475\textwidth}
        \centering
        \includegraphics[height=4.5cm]{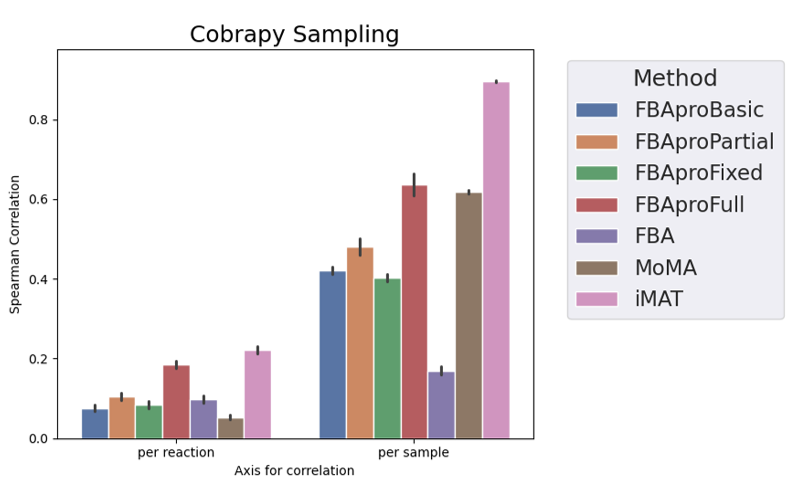}
        \caption{Spearman correlations between steady-state fluxes and method predictions, data sampled using cobrapy's flux sampling algorithm.}
        \label{fig:synthetic_correlations_b}
    \end{subfigure}
    \caption{Method performance on the synthetic data inference experiments. Spearman correlations between method predictions given partially noisy and missing data, measured on noisy reactions for each sample and each reaction. Error bars represent a 95\% confidence interval.}
    \label{fig:synthetic_correlations}
\end{figure}

Figure \ref{fig:synthetic_correlations_a} shows the average Spearman correlations for each method, computed both across reactions and samples. In this setting, iMAT returned constant zero predictions. The FBApro variants outperformed the other benchmark methods, with FBAproFull achieving the best correlations on both the reaction and sample axes. All FBApro variants performed better than MoMA and FBA, demonstrating that FBApro can recover hidden flux structure more accurately than existing approaches when the synthetic data are generated from the steady-state space. 

To assess the sensitivity of these results to the details of the synthetic data generation procedure, we repeated the experiment using steady-state flux samples generated through cobrapy's sampling method, which produces samples constrained by the model's reaction bounds. These results are shown in Figure \ref{fig:synthetic_correlations_b}. Overall, correlations are substantially lower for all methods under this sampling scheme, particularly when comparing predictions for each reaction across samples. In this setting, iMAT achieves the highest correlations while most FBApro variants continue to outperform FBA, and FBAproFull remains competitive with MoMA.

Taken together, these experiments show that FBApro achieves state-of-the-art flux recovery performance, although the top-performing method can depend on the sampling procedure used to generate synthetic fluxes. Across both sampling schemes, FBApro remains among the stronger performing methods, and in the first setting it is the clear best performer. These accuracy results, together with the amortized timing advantages described above, indicate that FBApro provides a favorable combination of predictive performance and computational efficiency. 

\subsection{Predicting Cancer Cell Line Exchange Rates}

To evaluate FBApro on real data, we used metabolite uptake and secretion rates measured for 52 cell-lines from the NCI60 panel~\cite{jain2012}. These measurements were matched to 49 existing exchange reactions in the RECON1 metabolic model. We additionally matched 45 of these cell lines with GE data from CCLE~\cite{barretina2012cancer}, reported in TPM units. The GE data was normalized and mapped to the model reactions to produce reaction activities in the range $[0, 1]$. To obtain reaction-specific flux ranges, we ran FVA on the RECON1 model using 90\% optimality and loopless constraints; see Appendix Section \ref{apxsec:fva} for more details. The reaction activities were then mapped to GE-derived flux estimates via a linear mapping for each reaction from $[0, 1]$ to the range between its lower and upper bound. For reactions with no associated data, we used the mean between the lower and upper bound of the reactions as an estimate. See Appendix Section \ref{apxsec:ge_processing} for more details on GE preprocessing. 

We performed two experiments on the data. In the leave-one-out experiment, we use only the exchange rate measurements as reference flux data, filling the rest with the mean of lower and upper reaction bounds from FVA. For each sample, we mask one exchange reaction at a time, replacing its value with the mean of bounds as well. We then apply all methods to predict its flux value, with the unmasked exchange rates considered as fixed values, the masked as unknown, and the rest as noisy. The concatenated predictions across all masked reactions  form the predicted fluxes by each method for each sample. Note that this results in $2548=49\cdot 52$ applications of each method to the data, highlighting the importance of efficient methods for working with metabolic data. 

In the GE integration experiment, we used only the GE data for reference fluxes as input, and evaluated how well each method recovered the measured exchange rates. Reactions with associated GE data were treated as measured, and the rest as unknown. This setting is more challenging than the leave-one-out experiment because the input data and evaluation targets come from different molecular modalities, and are linked only indirectly through the metabolic model. For both experiments and methods taking bounds as input, we used a gap of |x| for a reaction with reference value $x$.

\begin{figure}[htbp]
    \centering
    \begin{subfigure}[b]{0.475\textwidth}
        \centering
        \includegraphics[height=4.5cm]{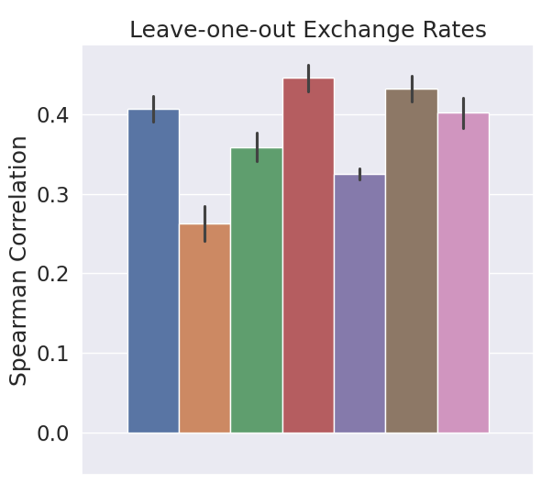}
        \caption{Spearman correlations per-sample between NCI60 exchange rate data and method predictions on masked exchanges.}
        \label{fig:real_correlations_a}
    \end{subfigure}
    \hfill
    \begin{subfigure}[b]{0.475\textwidth}
        \centering
        \includegraphics[height=4.5cm]{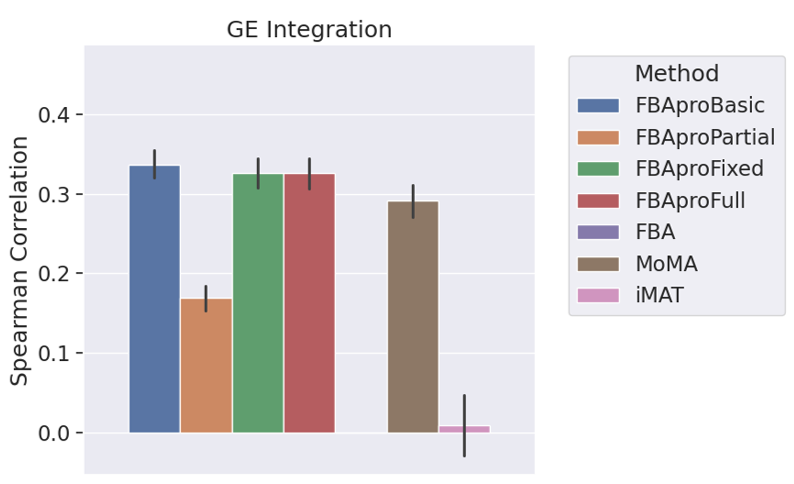}
        \caption{Spearman correlations per-sample between NCI60 exchange rate data and method predictions on GE mapped reference fluxes}
        \label{fig:real_correlations_b}
    \end{subfigure}
    \caption{Method performance on the real data inference experiments. Spearman correlations between method predictions and NCI60 exchange rate measurements, computed per sample. In the leave-one-out experiment, each exchange rate is masked in turn and predicted using the remaining measured exchange rates as input.  In the GE integration experiment, matched GE data is used to create a reference flux vector and method predictions are compared with measured exchange rate data. Error bars represent a 95\% confidence interval.}
    \label{fig:real_correlations}
\end{figure}

Figure \ref{fig:real_correlations} shows the Spearman correlations between predictions and measured exchange fluxes, computed separately per sample. In the leave-one-out experiment, where the input and target measurements come from the same data source, units and experimental setup, correlations are typically high. FBApro variants are competitive with the other benchmark methods, with FBAproFull performing the best. In the GE integration experiment, where the GE-derived reaction activity data is fundamentally different than the fluxes, most FBApro variants still achieve high correlations and outperform benchmark methods. In contrast, FBA predicts constant zero fluxes for all evaluated reactions, while iMAT predicts 89\% of the fluxes as zero. These results suggest that FBApro can effectively integrate heterogeneous real-world measurements and achieve strong predictive performance.

\section{Discussion}\label{sec:discussion}

In this work we present FBApro, a linear transformation solving the problem of finding a closest steady-state flux to a noisy, partial reference flux vector. We define the optimization problem solved by FBApro, as well as special cases of it, derive the expression for the corresponding linear transformations, and prove these are indeed optimal solutions. We demonstrate the usefulness of FBApro on an example toy model. We further demonstrate that FBApro is computationally scalable to large amounts of data, allowing a qualitative difference in the kinds of data and \emph{in vitro} experiments that can be utilized with it. FBApro also performs on par or better than existing methods in denoising and recovering missing data, as well as on predicting fluxes based on GE.

A major factor of the centrality of FBA in metabolic modeling is the ease with which it can be modified. Alongside its relative scalability, this resulted in a wide variety of methods in the field extending the optimization problem or utilizing FBA as a building block within their algorithm. We believe that FBApro, as another simple, flexible method, can also serve as another basic building block and inspire more diverse future methods.

Even without further modification, FBApro has a variety of potential use-cases: GE data integration, where GE of genes for a reaction are used to make initial estimates of fluxes; noise cancellation and data imputation, where noisy and/or missing flux measurements or estimates are mapped to steady-state fluxes; simulation of knockdowns, where wild-type fluxes are replaced with zeros for knocked down reactions; and even  simulating different metabolic objectives, by finding the closest steady-state vector to an indicator vector of the corresponding reactions.

An important consideration in using FBApro is that it does not enforce reaction bounds. This enables a substantially simpler framework, but also discards potentially useful information. In FBA and related methods, reaction bounds are often needed to ensure a finite solution; however, their values are typically unknown and frequently set to default values. In our validations, FBApro performs favorably without bounds. Furthermore, context-specific estimated bounds can be interpreted as estimated flux values with associated confidence ranges and incorporated as reference fluxes. Nevertheless, FBApro cannot enforce reaction directionality constraints, and this remains an important limitation of the approach. 

One motivation for replacing linear programming with a linear transformation is that the resulting closed-form differentiable function can easily fit within gradient-based learning frameworks. FBApro could improve  metabolic modeling in these frameworks, either as a concept bottleneck preventing overfitting, or as a mechanism for transfer learning by constraining an inner layer to have a structure interpretable as reaction fluxes. Integrating neural architectures with CBMM modeling assumptions is a promising avenue for future research \cite{faure2023neural, alghamdi2021graph}, and by mapping model outputs directly into the steady-state space, FBApro may provide a simple but powerful bridge between flexible neural architectures and mechanistically constrained metabolic models.

\newpage

\bibliography{references}

\newpage 
\appendix
\section{Background}
\subsection{Constraint-Based Metabolic Modeling and Flux Balance Analysis}
\label{apxsec:fba}

Let $S\in \mathbb{R}^{m\times r}$ be a stoichiometric matrix whose rows correspond to metabolites and columns to reactions. The value of $S_{i,j}$ represents the concentration (or amount) of metabolite $i$  (e.g. in micromolar units $\mu M$) produced (or consumed, if negative) by one unit of flux of reaction $j$, in inverse time unit (e.g. $1/sec$). Given a vector of flux rates $v\in \mathbb{R}^r$, the vector $Sv\in \mathbb{R}^m$ represents the rate of change in each metabolite. The steady-state assumption at the heart of CBMM is then $Sv=0$. 

FBA aims to simulate the fluxes in the cell (or tissue, organism, community) that $S$ represents, under the assumption that metabolic processes aim to maximize a linear objective in terms of reaction fluxes $c\in \mathbb{R}^r$. Typically, $c$ represents growth, by giving a positive coefficient to a reaction representing the consumption of different metabolites for the creation of biomass. The objective value for a flux vector $v$ is then $c^Tv$. Since the objective is linear, any $v$ with nonzero objective value can be arbitrarily scaled to achieve any objective value. Therefore, CBMM adds constraints to the fluxes of each reaction preventing arbitrary scaling, and potentially encoding real biological constraints of reaction direction, enzyme abundance or metabolite availability. Let $lb,ub\in \mathbb{R}^r$ be vectors representing the lower and upper bounds for reaction fluxes, respectively. FBA is the linear program solving for an optimal flux vector, that is 

\begin{align*}
max_v & &c^Tv\\
&\text{subject to} &Sv=0\\
&&  v\leq ub\\
&&  v\geq lb.
\end{align*}

\subsection{Flux Variability Analysis}\label{apxsec:fva}

Flux Variability Analysis (FVA) is a method used to automatically adjust the bounds of reactions in a model to represent their feasible flux ranges. Intuitively, every solution to FBA, for a non-degenerate model, has critical inequalities that are satisfied with an equality in an equation in that solution. However, an inequality may remain unsatisfied in any optimal (or near-optimal) solution, e.g. an upper bound on reactions involved in glycolysis may be higher than the maximal rate of uptake and production of glucose in the model. FVA aims to replace reaction bounds such that each one has a near-optimal solution in which it is critical. 

Formally, let $\alpha\in [0,1]$ be a fraction of optimality, and $c, S, lb, ub$ as in the formulation of FBA, and $x$ be the optimal value of FBA's objective. Then for every reaction $j$, FVA computes 

\begin{align*}
lb^*_j = min_v & & v_j\\
&\text{subject to} &Sv=0\\
&&c^Tv\geq \alpha x\\
&&  v\leq ub\\
&&  v\geq lb,
\end{align*}

and similarly $ub_j^*$ with maximization substituted for minimization. This amounts to two FBA-like linear programs run for each reaction in the model.

A common modification to FVA is to solve the aforementioned program with the added requirement of no flux loops, also known as futile cycles or thermodynamically infeasible cycles. These are typically defined as flux vectors with non-zero fluxes only for mass-preserving reactions, and no exchange or demand reactions. 

\subsection{Context-Specific CBMM Methods}\label{apxsec:imatmoma}

Many methods have been developed to simulate the metabolism of systems under a specific context. Such a context can be a gene knockout strain, a specific growth medium, measured fluxes or gene expression. We cover two such methods here. Minimization of Metabolic Adjustment (MoMA) receives as input a metabolic model with stoichiometry $S$ and bounds $lb, ub$, and a reference flux $v$. While the method is generic, the original and still standard implementation is in the context of a knockout strain, where $v$ is taken from a wild-type model, whereas $S$ represents some subset of the original model's reactions. MoMA is then the quadratic program

\begin{align*}
min_x & &\parallel v - x \parallel^2_2\\
&\text{subject to} &Sx=0\\
&&  x\leq ub\\
&&  x\geq lb.
\end{align*}

Integrative Metabolic Analysis Tool (iMAT) is designed to simulate reaction fluxes given GE input. Assume the GE has been mapped to reactions resulting in a reference vector $v$ as in MoMA, albeit with nonnegative values. However, given the inherent noise and loose relationship between genes and reaction fluxes, iMAT interprets the reference vector as indicating only which reactions are likely active and which ones aren't. Let $a, b\in [0,1]$ be fixed thresholds for interpreting active and inactive indications. Let $x$ be the percentile ranks of the values of $v$, such that $x_i\in [0,1]$ represents the fraction of reactions with lower or equal values, or $\left|\{j|v_j\leq v_i\}\right|/r$. Then the reactions of the models are partitioned to $L=\{i|x_i< a\}$, $H=\{i|x_i> b\}$, $M$ otherwise. Presented verbally and simplified here, iMAT is the mixed-integer program maximizing the number of reactions in $H$ with non-trivial flux, and the number of reactions in $L$ with zero flux, conforming to steady-state assumptions. For our experiments, we used $a=1/3, b=2/3$.

\section{FBAproBasic Proof}
\label{apxsec:basic_proof}

Here we prove Theorem \ref{thm:basic}, representing $FBAproBasic(S, v)$ as a projection to the space $ker(S)=col(A)$. This is a standard orthogonal projection, and while a reader familiar with projections using pseudoinverses can safely skip this proof, it is our hope that some readers will find it instructive and preliminary to the proofs for the more general variants of FBApro.

\begin{proof}[Proof of Theorem \ref{thm:basic}]
Let $S^+$ be the pseudoinverse of $S$, following the real-value versions of the Moore-Penrose conditions, and computable using SVD:

\begin{itemize}
\item $SS^+S=S$
\item $S^+SS^+=S^+$
\item $(SS^+)^T=SS^+$
\item $(S^+S)^T=S^+S$
\end{itemize}

We want to show that for every $v\in \mathbb{R}^r$, $Q=I-S^+S$ is such that $Qv$ is the closest vector to $v$ in $ker(S)$. First, observe that 

\[SQv=S(I-S^+S)v=(S-SS^+S)v=(S-S)v=0,\]
and so $Qv\in ker(S)$. To show it minimizes distance to $v$ among all steady-state vectors $x\in ker(S)$, note that 

\[
\lVert v-x\rVert_2^2=\lVert \left((v-Qv)+(Qv-x)\right)\rVert_2^2=\lVert v-Qv\rVert_2^2+\lVert Qv-x\rVert_2^2+2\langle v-Qv,Qv-x\rangle.
\]
To show that $\lVert v-x\rVert_2^2\geq \lVert v-Qv\rVert_2^2$ it suffices to show that $\langle v-Qv,Qv-x\rangle=0$. Then 

\begin{align*}
\langle v-Qv,Qv-x\rangle=(v-Qv)^T(Qv-x)=v^T(I-Q)^T(Qv-x)=\\
v^T(S^+S)^T(v-S^+Sv-x)=v^T(S^+S)(v-S^+Sv-x)=\\
v^T(S^+Sv-S^+SS^+Sv-S^+Sx)=-v^TS^+(Sx)=0.
\end{align*}

To represent FBAproBasic using $A$, first recall that its columns are orthonormal, and therefore $A^TA=I$. It is thus easy to verify that the Moore-Penrose conditions hold with $A^+=A^T$. Since $col(A)=ker(S)$, $AA^Tv=A(A^Tv)\in col(A)=ker(S)$. For distance minimization, let $x\in ker(S)$, then $\exists y.x=Ay$, and

\begin{align*}
    \langle v-AA^Tv,AA^Tv-x\rangle =v^T(I-AA^T)^T(AA^Tv-Ay)=v^T(I-AA^T)A(A^Tv-y)=\\
    v^T(A-AA^TA)(A^Tv-y)=0.
\end{align*}
\end{proof}

\section{FBAproFixed Basic Linear Algebra Proof}
\label{apxsec:fixed_proof}

Recall the use of the pseudoinverse for solving systems of linear equations: if the system of linear equations $Tx=y$ admits a solution, then $x=T^+y$ is one. As a note, if multiple solutions exist, $T^+y$ has minimal norm among them, and if no solutions exist, $T^+y$ minimizes the error $\lVert Tx-y\rVert_2^2$, leading to its use in least-square regression problems. While not part of our proofs, this suggests that FBApro and FBAproFixed are expected to return an approximate solution even when an exact one agreeing with $v$ on $H$ may not exist. Given the use of the pseudoinverse for solving linear equations, we can prove Theorem \ref{thm:fixed1}:
 \begin{proof}[Proof of Theorem \ref{thm:fixed1}]

Consider the constraints $Sx=0$ and  $x_H=v_H$, and assume a solution exists. Treating $Sx$ as a combination of the columns of $S$, separate them to the columns corresponding to $H$ and $H^c$ and then substitute $v_H$ for $x_H$, yielding 

\begin{align*}
(S_{[m],H})x_H+(S_{[m],H^c})x_{H^c}&=0\\
(S_{[m],H^c})x_{H^c}&=-(S_{[m],H})v_H.
\end{align*}

Denote $Q_0\in \mathbb{R}^{|H^c|\times|H|}$ defined as $Q_0=-(S_{[m],H^c})^+(S_{[m],H})$. Then, if the system has any solutions, a particular solution is admitted by
\[
x'_{H^c}=Q_0v_H=-(S_{[m],H^c})^+(S_{[m],H})v_H.
\]
Since we are interested in a solution minimizing distance to $v$, we use this expression for a solution to express the affine space defined by the previous system of non-homogeneous equations:

\[
U=ker(S_{[m],H^c})+x'_{H^c}
\]
Any vector $x$ with $x_H=v_H$ and $x_{H^c}\in U$ satisfies the original constraints, and has 

\[
\lVert v-x \rVert_2^2=\lVert v_{H^c}-x_{H^c} \rVert_2^2,
\]
and so fixing $v_H$ and projecting $v_{H^c}$ to the closest point in $U$ solves Problem \ref{opt:fixed}. For any affine space $T+u$, a vector  $t+u$ in the space and a vector $w$, we have that 

\[
\lVert w-(t+u) \rVert_2^2=\lVert (w-u)-t \rVert_2^2
\]
and so an orthogonal projection minimizing distance to $T+u$, given an orthogonal projection matrix $Q_T$ to $T$, is of the form $w\rightarrow u+Q_T(w-u)$. For the linear space corresponding to $U$, following Theorem \ref{thm:basic}, the corresponding orthogonal projection is $Q_U\in \mathbb{R}^{|H^c|\times |H^c|}$ with

\[
Q_U=I_{|H^c|}-(S_{[m],H^c})^+S_{[m],H^c}.
\]

Note that 

\[
Q_UQ_0=-(S_{[m],H^c})^+S_{[m],H} + (S_{[m],H^c})^+S_{[m],H^c}(S_{[m],H^c})^+S_{[m],H}=0.
\]

Define $Q\in \mathbb{R}^{r\times r}$ such that

\begin{align*}
    &Q_{H,H}&=&I_{|H|}&&\\
    &Q_{H, H^c}&=&0&&\\
    &Q_{H^c,H}&=&-(S_{[m],H^c})^+(S_{[m],H})&=&Q_0\\
    &Q_{H^c,H^c}&=&\left(I_{|H^c|}-(S_{[m],H^c})^+S_{[m],H^c}\right)&=&Q_U.
\end{align*}

Then $Qv$ fixes $v_H$ as 

\[
(Qv)_H=(Q_{H,[r]})v=(Q_{H,H})v_H+(Q_{H,H^c})v_{H^c}=(I_{|H|}) v_H=v_H,
\]

and projects $v_{H^c}$ to $U$ as

\begin{align*}
(Qv)_{H^c}=(Q_{H^c,[r]})v&=(Q_{H^c,H})v_H+(Q_{H^c,H^c})v_{H^c}=\\&=Q_0v_H+Q_Uv_{H^c}+0=\\
&=x'_{H^c}+Q_U v_{H^c}-(Q_UQ_0)v_H=x'_{H^c}+Q_U(v_{H^c}-x'_{H^c}),
\end{align*}
as needed.
\end{proof}

\section{Processing and Mapping Gene Expression to Reference Fluxes}\label{apxsec:ge_processing}

GE datasets are processed as follows: Genes with less than 15\% non-zero values across samples are dropped. We then apply a quantile transformation to the expression data: denote the number of genes as $g$. We first create a reference distribution by taking, for every rank $i\leq g$, the mean expression value of the $i$-th smallest value in each sample. We then map the values of each sample to this distribution, preserving their ranking. The entire GE matrix is then shifted and scaled so that the values are in the range $[0, 1]$, to represent relative activity.

In order to apply any method to GE data, it must be mapped from the gene space to the reaction space of the model. We follow a typical process for the mapping, although the implementation details are not standardized in the literature \cite{metabo13010126}. We use the model's Gene-Protein-Reaction (GPR) association, which associates each reaction with a logical expression with gene ids as variables.  In the case of multiple transcript variants in the model, this information is ignored for matching with GE data. The logical operators AND and OR are replaced with geometric mean and arithmetic mean, respectively, and the GE values are substituted for the variables to result in raw reaction activities. Genes not in the data are ignored, so means are taken over the rest. Note that this preserves the range of the GE values. 

Let $lb, ub$ denote the lower and upper bounds of a reaction, and $x$ denote its activity in a sample. For a forward reaction, with $lb, ub\geq 0$, the corresponding reference flux is $lb + x(ub-lb)$, such that an activity of 0 corresponds to $lb$ and 1 to $ub$. For backward reactions with $lb,ub\leq 0$, the roles of $lb,ub$ are reversed in the mapping. For reversible reactions with $lb\leq 0\leq ub$, no value is assigned from activity. These, as well as reactions without a GPR, or without any matching genes in the data, are assigned a reference flux of $(lb+ub)/2$.  Note that after running FVA only a small fraction of the reactions remain bidirectional (323/3418).

\begin{figure}[htbp]
    \centering
    \begin{subfigure}[b]{0.475\textwidth}
        \centering
        \includegraphics[height=4.5cm]{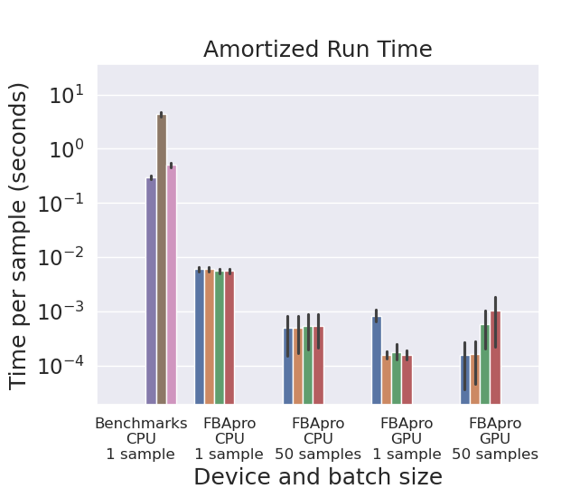}
        \caption{Running time for each method on a single sample, when run on CPU or GPU, and either on 1 sample or amortized over 50 samples. For FBApro, setup time is not included in the running time.}
        \label{fig:timing_a}
    \end{subfigure}
    \hfill
    \begin{subfigure}[b]{0.475\textwidth}
        \centering
        \includegraphics[height=4.5cm]{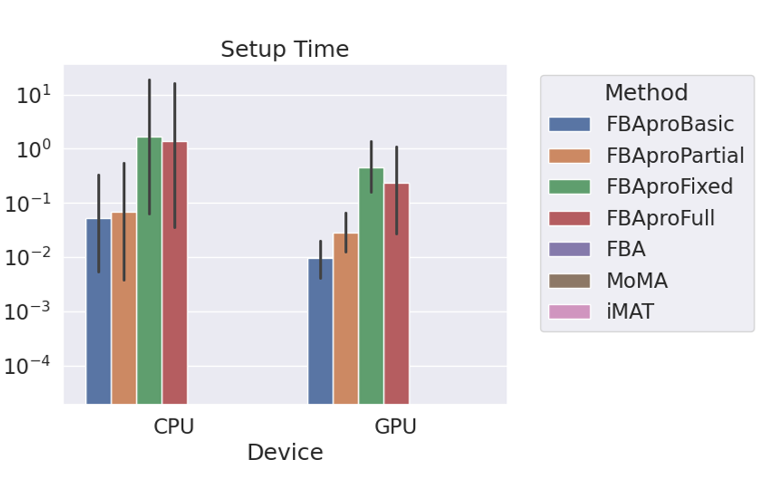}
        \caption{Setup time for FBApro variants on CPU and GPU devices.}
        \label{fig:timing_b}
    \end{subfigure}
    \caption{Timing results, in seconds, on a log scale, averaged over five models and 50 samples per model. For FBApro variants, runtime is measured on both CPU and GPU, and on the inputs given separately or as a single matrix. Error bars represent a 95\% confidence interval.}
    \label{fig:apx_timing}
\end{figure}

\end{document}